\documentclass[aps,prl,10pt,twocolumn]{revtex4-1}
\usepackage{amsmath,amssymb}
\usepackage{graphicx}
\usepackage[dvipsnames]{xcolor}
\usepackage{comment}
\usepackage[overload]{empheq}
\usepackage{upgreek}
\usepackage{lmodern,pdftexcmds}
\usepackage[sort&compress]{natbib}
\usepackage{cases}

\usepackage[normalem]{ulem}
\usepackage{xcolor}

\renewcommand{\bm}[1]{\boldsymbol{\mathbf{#1}}}

\begin{document}

\title{Universal Persistent Brownian Motions in Confluent Tissues}

\author{Alessandro Rizzi}
\affiliation{Institute of Mechanical Engineering, \'{E}cole Polytechnique F\'{e}d\'{e}rale de Lausanne (EPFL)}

\author{Sangwoo Kim*}
\affiliation{Institute of Mechanical Engineering, \'{E}cole Polytechnique F\'{e}d\'{e}rale de Lausanne (EPFL)}

\date{\today}

\begin{abstract}
% Need to write at the end

Biological tissues are active materials whose non-equilibrium dynamics emerge from distinct cellular force-generating mechanisms. Using a two-dimensional active foam model, we compare the effects of traction forces and junctional tension fluctuations on confluent tissue dynamics. While these two modes of activity produce qualitatively different cell shapes, rearrangement statistics, and spatiotemporal correlations in fluid states, we find that the long-time cellular motion universally converges to persistent Brownian dynamics. This universal feature contrasts with the non-universal correlations between cell geometry, rearrangement rate, and fluidity, which depend sensitively on the underlying modes of active force. Our results demonstrate that persistent Brownian motion provides a minimal framework for describing tissue dynamics, while distinct active forces leave identifiable structural and dynamical signatures, thereby enabling inference of the dominant active force in fluid state tissues.

\end{abstract}

\maketitle
Biological tissues are a prime example of active matter that constantly consumes energy at cellular and subcellular scales, driving dynamical changes in tissue architecture and mechanical states. This dynamic regulation plays a key role in many biological processes, including embryonic development, homeostasis, and disease progression \cite{guillot2013mechanics,heisenberg2013forces}. Although signaling pathways and gene expressions have been central in understanding these processes, recent studies highlight that modulating cellular mechanical properties is essential for robust control of tissue mechanical states.

In particular, accumulating studies show that biological tissues can induce phase transition behaviors between solid and fluid states through distinct mechanisms \cite{lawson2021jamming,hannezo2022rigidity,lenne2022sculpting,mao2024mechanical}. An increase in cell packing density within an epithelial monolayer exhibits glassy dynamics and caging behaviors \cite{angelini2011glass}, reminiscent of jamming and glass transition in soft materials \cite{o2003jamming,van2009jamming}. A confluent epithelial monolayer can also undergo phase transition by reaching a zero-tension state \cite{park2015unjamming}. More recent studies show that phase transitions, including jamming, glass, and percolation transitions, play a key role in embryonic development across various organisms \cite{mongera2018fluid,petridou2019fluidization, petridou2021rigidity, kim2024nuclear,huycke2024patterning}. Complementing experimental observations, theoretical and computational models elucidate different modes of phase transitions, providing the physical basis for explaining such behaviors \cite{staple2010mechanics,bi2015density,kim2015cell,bi2016motility,krajnc2020solid,kim2021embryonic,kim2024nuclear,li2025fluidization}.

While it is well established that increasing the magnitude of active forces leads to activity-induced phase transitions~\cite{bi2016motility,malinverno2017endocytic,kim2021embryonic}, how distinct modes of activity shape structural and dynamical features in non-equilibrium fluid states remains unclear. In this paper, we examine two well-known modes of active forces in densely packed monolayer systems, namely traction forces and junctional tension fluctuations, focusing on regimes in which one mode is dominant over the other, to identify similarities and differences in their emergent properties. 

\begin{figure}[b]
    \centering
    \includegraphics[width=1\linewidth]{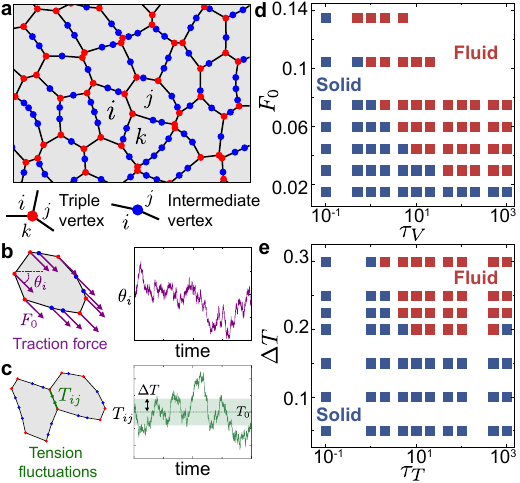}
    \caption{Modeling framework and phase transition behaviors. (a) Representative tissue snapshot, showing distinct types of vertices. (b) Traction force as a self-propelling force, whose polarity $\theta_i$ diffuses over time. (c) Edge tension $T_{ij}$ fluctuates around a fixed point $T_0$. Phase diagram of (d) traction force, and (e) tension fluctuations, based on long-time MSRD.}
    \label{fig:1}
\end{figure}
% To investigate the dynamics of confluent multicellular tissues, 
To that end, we employ a two-dimensional active foam model~\cite{kim2021embryonic} under the confluent condition, i.e., with no spaces between cells. The geometry of individual cells follows the vertex model framework, in which vertices at interfaces describe cell contours. In addition to the physical triple vertices at the tricellular junctions, intermediate vertices are introduced to capture complex cell shapes and junctional curvature (Fig.~\ref{fig:1}(a)). The tissue structure and dynamics follow from the vertex dynamics, described by an overdamped Langevin equation: 
\begin{equation}
    \eta_{R} \frac{\mathrm{d}\bm{R}_\alpha}{\mathrm{d}t} = \sum_{<i,j>} \biggl[ \bm{T}_{ij} \Theta(T_{ij}) + \bm{N}_{ij}   \biggr] + \sum_{k} \bm{F}_k.
    \label{eq:eom_R}
\end{equation} 
Here, ${\bm{R}}_\alpha$ denotes the position of vertex $\alpha$, and $\mathbf{T}_{ij}$ and $\mathbf{N}_{ij}$ are the tangential and normal forces at the junction between cell $i$ and $j$ associated with vertex $\alpha$, and $\bm{F}_{k} = F_0\left(\cos\theta_k,\sin\theta_k\right)$ is the traction force of cell $k$ acting on vertex $\alpha$. The effective force comprises three tangential, three normal, and three traction forces at a triple vertex, and two of each at an intermediate vertex (Fig. S7). The Heaviside step function $\Theta(T_{ij})$ prevents non-physical negative tension. The tangential force arises from the combined effects of cortical tension and cell-cell adhesion, setting a tension scale $T_0$, while the normal force results from osmotic balance between cells. Note that the tangential and normal forces from standard vertex model energy functionals with area and perimeter elasticity can be mapped directly onto our formalism (see Supplementary Information, Section 2.3 and Fig. S8).

We consider two dominant modes of activity in isolation: traction forces and tension fluctuations. Cells generate traction force through focal adhesion, cytoskeletal reorganization, and actomyosin contractility~\cite{brugues2014forces,alert2020physical}, which we model as a self-propulsion $F_0$ with polarity undergoing rotational diffusion as $\tau_V \mathrm{d}\theta_i/\mathrm{d}t = \xi$, where $\tau_V$ is the persistent timescale and $\xi$ is the standard Gaussian noise of null mean and unit standard deviation, as in an active Brownian particle~\cite{romanczuk2012active} (Fig.~\ref{fig:1}(b)).
Although this force acts on the cell center, it can be equivalently represented as the same force acting on all vertices independent of intermediate vertices resolution, resulting in a constant speed $F_0/\eta_R$ of an isolated cell, similar to traction implementation in previous studies \cite{sussman2017cellgpu,sadhukhan2024motility,de2025epithelial}.

Tension fluctuations arise from the dynamical regulation of the actomyosin cortex and cell-cell adhesion. We model these stochastic fluctuations using an Ornstein-Uhlenbeck process \cite{curran2017myosin,krajnc2020solid,kim2021embryonic}, in which the instantaneous tension fluctuates around the mean level $T_0$ by a fluctuation magnitude $\Delta T$ and a relaxation timescale $\tau_T$ (Fig.~\ref{fig:1}(c)).

\begin{equation}
\tau_T \frac{\mathrm{d}T_{ij}}{\mathrm{d}t} = - (T_{ij} - T_0) + \Delta T \sqrt{2 \tau_T} \xi.
\end{equation}

Both activities are characterized by two parameters, $F_0$ and $\tau_V$ for traction, and $\Delta T$ and  $\tau_T$ for tension fluctuations. We will investigate these parameter pairs separately without mixing (i.e. $\Delta T=0$ for traction and $F_0=0$ for tension fluctuations), while keeping other non-dimensional parameters fixed at biologically relevant scales (see SI, Section 2) Even for mixed cases, we confirm that our main results remain quantitatively unchanged when the characteristic scale of dominant active force is much larger that of the others (see SI, Section 4 and Fig. S9).

We first investigate how the tissue mechanical states shift from solid to fluid under each mode of activity by computing the mean-squared relative displacement (MSRD); $MSRD(t)=\left<\left(\bm{r}_{ij}(t+t_0)-\bm{r}_{ij}(t_0)\right)^2\right>$, where $\bm{r}_{ij}$ is the displacement vector between cells $i$ and $j$, which were neighbors at time $t_0$. Both active forces drive a transition from caging to subdiffusive and eventually diffusive behavior as activity increases, with superdiffusive exponents observed for traction forces at large $F_0$ and $\tau_V$(Fig.~S1(a)-(d)). 
% Because most MSRD curves are not diffusive even at long times in simulations, a diffusion coefficient cannot be reliably computed. Instead, 
We evaluate the MSRD at a fixed time, $t^{*}=2500\tau_R$ (Fig.~S1(e)(f)) and classify states as fluid when $MSRD(t^*)>1$ (Fig.~\ref{fig:1}(d)(e)), which was shown to effectively distinguish solid and fluid states in non-equilibrium systems \cite{mongera2018fluid,kim2021embryonic}. Note that the specific choice of $t^{*}$ only induces quantitative shifts in phase boundaries if $t^{*}$ is larger than all other characteristic timescales in the system (Fig.~S2). 

\begin{figure}[b]
    \centering
    \includegraphics[width=\linewidth]{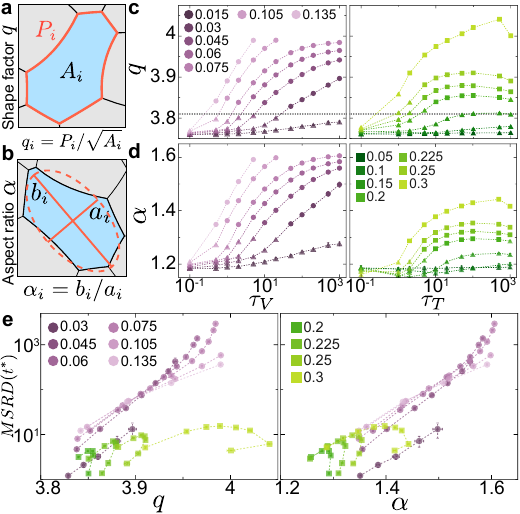}
    \caption{Cell geometry in fluid states. Schematic representation of (a) shape factor $q$ and (b) aspect ratio $\alpha$. Cell anisotropy changes for traction (left) and tension fluctuations (right) in terms of (c) shape factor and (d) aspect ratio. The dashed line in (c) indicates $q_c = 3.81$. Triangles, circles, and squares respectively represent solid,  fluid(traction), and fluid(tension fluctuations) states, as defined in Fig.~\ref{fig:1}(d)(e). (e) Correlations between shape factor, aspect ratio, and MSRD, exhibiting non-universal features.}
    \label{fig:2}
\end{figure}

Increasing the magnitude of activity, $F_0$ or $\Delta T$, drives a transition from solid to fluid states, but the effect of the persistent timescale differs for the two modes. Increasing $\tau_V$ for traction monotonically fluidizes the system, whereas increasing $\tau_T$ for tension fluctuations induces a non-monotonic behavior in $MSRD(t^*)$, with maximal fluidity at intermediate persistence (Fig.~\ref{fig:1}(d)(e), Fig.~S1(e)(f)). This contrast reflects intrinsic differences between the two modes. A large $\tau_V$ prolongs directed forces, enabling rearrangement even at lower $F_0$. In contrast, both small and large $\tau_T$ hinder fluidization, as short $\tau_T$ does not provide sufficient persistence of junctional shortening, while long $\tau_T$ causes divergence between shortening and relaxation timescales. The resulting phase diagrams agree with previous studies \cite{bi2016motility,barton2017active,kim2021embryonic,yamamoto2022non}, confirming that our model captures known non-equilibrium phase transitions driven by active forces. In the following, we will focus on fluid states to investigate how cell geometry, cellular rearrangements, and spatiotemporal correlations depend on the activity mode.

\begin{figure}[b]
    \centering
    \includegraphics[width=\linewidth]{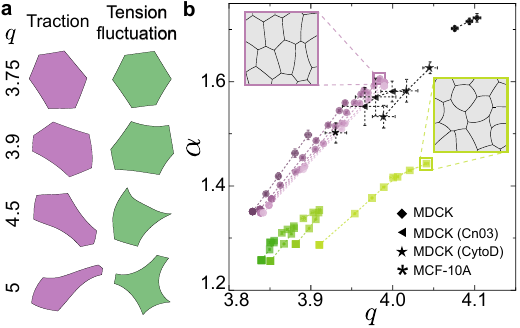}
    \caption{Cell geometry in fluid states. (a) Representative cell shapes for distinct $q$. (b) Correlations between $q$ and $\alpha$ for fluid states with experimental data~\cite{saraswathibhatla2020spatiotemporal,malinverno2017endocytic} shown as black markers. Error bars indicate the standard error of the mean. The insets illustrate the different morphologies at large $q$. }
    \label{fig:3}
\end{figure}

% We examine how cell geometry in fluid states varies with different activities,
We first examine cell anisotropy, a key geometric signature of fluid states~\cite{bi2015density,kim2021embryonic}, using two measures, the average shape factor $q=\left<q_i\right>$ and aspect ratio $\alpha=\left<\alpha_i\right>$. For each cell, the shape factor is defined as the ratio of the perimeter to the square root of area, $q_i=P_i/\sqrt{A_i}$ (Fig.~\ref{fig:2}(a)), and the aspect ratio is obtained by fitting an ellipse of major and minor axes $b_i$ and $a_i$ that preserves the eigenvalues of the second moment area, $\alpha_i=b_i/a_i$ (Fig.~\ref{fig:2}(b)).

For both traction and tension fluctuations, greater cell anisotropy correlates with increased motility and fluidity (Fig.~\ref{fig:2}(c-e)). For traction, the previously reported critical shape factor, $q_c\approx3.81$~\cite{bi2015density,bi2016motility}, accurately predicts the solid-fluid boundary identified by $MSRD(t^*)$. However, the phase boundary prediction is less reliable for tension fluctuations, with solid states of $q>q_c$ at intermediate $\tau_T$(Fig.~\ref{fig:2}(c)). The aspect ratio shows similar correlations, but its range is significantly narrower under tension fluctuations than under traction (Fig.~\ref{fig:2}(d)). These differences manifest as non-universal correlations between $q$ and $MSRD(t^*)$, and between $\alpha$ and $MSRD(t^*)$ (Fig.~\ref{fig:2}(e)), implying that cell shape alone cannot robustly infer the level of fluidity, particularly in systems dominated by distinct active forces.

How cells deform from the isotropic shape also strongly depends on the type of activity. We compare representative cell geometries at fixed $q$ under traction and tension fluctuations (Fig.~\ref{fig:3}(a)). At high $q$, cells under traction exhibit uniaxial elongation, whereas cells under tension fluctuations develop highly curved long edges. While directed traction forces induce elongation and nematic order (Fig.~S3), increasing tension fluctuations broaden the tension distribution, producing long, low-tension junctions with high curvature due to the Young-Laplace equation. Such lax junctions increase $q$ without increasing cell elongation, yielding smaller aspect ratios at the same $q$ (Fig.~\ref{fig:3}(b)).

To validate this, we analyze MDCK~\cite{saraswathibhatla2020spatiotemporal} and MCF-10A~\cite{malinverno2017endocytic} monolayers in fluid states, where dominant activity is traction force (see SI, Section 5 and Fig. S10). MCF-10A and Cn03-treated MDCK cells match our simulations, while cytochalasin D-treated MDCK cells show reduced aspect ratios consistent with suppressed motility. Low-density MDCK cells fall outside of our simulation range but follow the same trend , confirming consistency with our results. 

A similar non-universal feature appears in human bronchial epithelial layers undergoing either compression-induced unjamming or partial epithelial-to-mesenchymal transition (pEMT)~\cite{mitchel2020primary}, where lax junctions in the pEMT arise from reduced mean tension rather than fluctuations. Our results show that the same non-universal feature can emerge in non-equilibrium regimes. 

\begin{figure}[t]
    \centering
    \includegraphics[width=\linewidth]{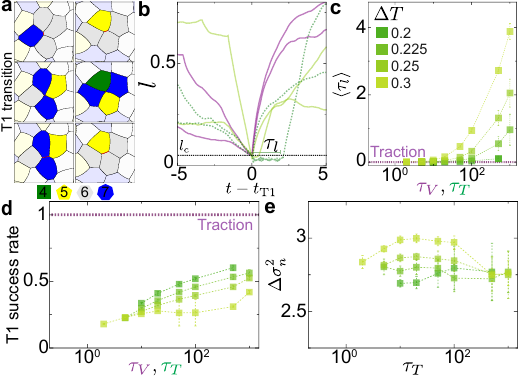}
        \caption{T1 dynamics. (a) Schematics of a successful (left) and an unsuccessful (right) T1 transition. (b) Representative edge length evolution during a T1 transition for traction (purple) and tension fluctuations (green). Solid and dashed lines represent successful and unsuccessful T1 events, respectively. The gray line marks $l_c$. (c) Mean stalling time, and (d) T1 success rate for tension fluctuations. (e) Mean expected change in defect density for unsuccessful T1 transitions.}
    \label{fig:4}
\end{figure}

We next analyze cellular rearrangements, an essential feature of fluidity. In confluent layers, these occur through T1 transitions, i.e., the loss of existing contacts and the formation of new junctions \cite{walck2014cell,bi2014energy,krajnc2018fluidization,kim2018universal}(Fig.~\ref{fig:4}(a)). For both activities, higher fluidity corresponds to a higher T1 rate, $\theta_\text{T1}$ (Fig.~S4). To characterize T1 dynamics, we track the length of each edge and identify $t=t_\text{T1}$ when its length falls below the critical threshold $l_c$ (Fig.~\ref{fig:4}(b), SI). Under traction, all short edges elongate immediately after the transition with zero stalling time $\tau_l=0$. In contrast, under tension fluctuations, a fraction of edges exhibit finite stalling times, with both the stalled fraction (Fig.~S5(b)(c)) and mean stalling time (Fig.~\ref{fig:4}(c), Fig.~S5(d)) increasing with $\tau_T$.

Beyond finite $\tau_l$, the T1 success rate also differs qualitatively. We define a successful T1 as one that produces a net topology change (Fig.~\ref{fig:4}(a)). Under traction, the success rate is 1, whereas tension fluctuations produce a substantial fraction of unsuccessful events, reverting to the original topology (Fig.~\ref{fig:4}(a)(d)). This difference stems from force persistence; traction provides a persistent force throughout the rearrangement, whereas junctional shortening due to tension fluctuations loses directional memory after the transition. This result is consistent with previous studies with mechanical feedback~\cite{perez2023tension}.  

An energy argument between adjacent local energy minima explains unsuccessful T1 transitions. Although the system is far from equilibrium, the underlying local energy minima and their energy level can be inferred from the defect density, $\sigma_n^2=\left<(n_i-6)^2\right>$, where $n_i$ is the number of neighbors of cell $i$. In two-dimensional monodisperse systems, higher $\sigma_n^2$ corresponds to higher energy~\cite{kim2019simple}. For unsuccessful T1 transitions, we compute the expected change in $\sigma_n^2$ assuming successful ones, $\Delta\sigma_n^2$, and find that they are biased toward energy-increasing configurations (Fig.~\ref{fig:4}(e)). In contrast, successful T1 transitions yield $\left|\Delta\sigma_n^2\right| \ll 1$, as required for steady states. Thus, energy landscape and local energy barriers strongly influence T1 dynamics under tension fluctuations, but not under traction forces. 

\begin{figure}
    \centering
    \includegraphics[width=\linewidth]{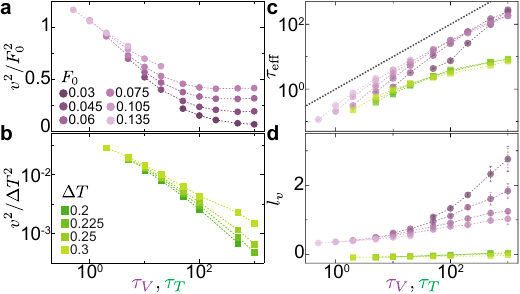}
        \caption{Spatio-temporal correlations in cell dynamics. Normalized mean squared velocity for (a) traction (linear scale) and (b) tension fluctuations (log scale). (c) Effective persistence $\tau_\text{eff}$ and (d) correlation length $l_v$. The dashed line in (c) represents unity.}
    \label{fig:5}
\end{figure}

Changes in the persistence of activity also affect cell velocity in distinct ways. Increasing persistence reduces the mean squared velocity, $v^2 = \left< v_i^2 \right>$, in both cases, but traction reaches a constant speed, while tension fluctuations slow down to zero (Fig.~\ref{fig:5}(a)(b)). The velocity plateau for traction arises from the balance between self-propulsion $F_0$ and passive force set by $T_0$ at large $\tau_V$, similar to dense self-propelled particles~\cite{szamel2024extremely}. On the other hand, large $\tau_T$ slows down junctional dynamics, solidifying tissue and reducing cell speed.

The spatial and temporal velocity correlations also show qualitative differences between the two activity modes. We compute temporal autocorrelation and spatial correlation functions, extracting an effective persistence time, $\tau_\text{eff}$, and a velocity correlation length, $l_v$ (see SI, Section 3). Increasing $\tau_V$ or $\tau_T$ enhances velocity persistence, though the effect is much weaker for tension fluctuations (Fig.~\ref{fig:5}(c), Fig.~S6(a)(b)). While the inherent persistence for traction directly governs cell velocity, the persistence in tension fluctuations couples weakly to cell motion, making it less effective in increasing $\tau_\text{eff}$. Furthermore, due to the sharp decrease in velocity, the persistent length of the single cell trajectory under tension fluctuations remains much smaller than a cell length even for large $\tau_\text{eff}$ (Fig.~S6(c)). Under traction, collective motion spontaneously emerges with increasing $\tau_V$, consistent with previous studies~\cite{henkes2020dense,caprini2020spontaneous,keta2022disordered,li2024relaxation}, while tension fluctuations consistently give $l_v\ll1$, indicating the absence of collective motion (Fig.~\ref{fig:5}(d), Fig.~S6(d)-(g), Supplementary movies 1-4). 
% These results demonstrate that the accessible ranges of $\tau_\text{eff}$ and $l_v$ are clearly separated between traction and tension fluctuations, implying that specific non-equilibrium dynamical features require distinct modes of activity. 
%keta2025long,sadhukhan2024motility

\begin{figure}[b]
    \centering
    \includegraphics[width=\linewidth]{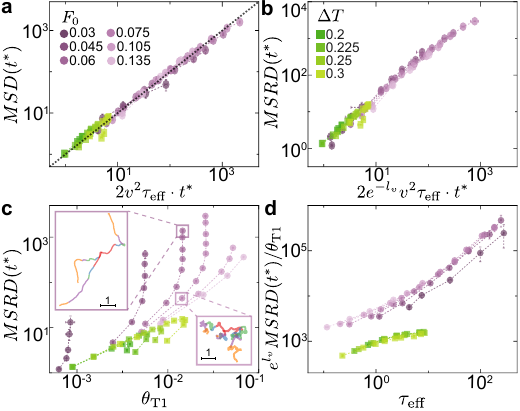}
        \caption{Universal and non-universal correlations with cell movement. Long-time (a) MSD and (b) MSRD can be predicted by persistent Brownian motion. (c) The MSRD is not proportional to the T1 rate. The two insets show three cell trajectories over five T1 transitions with color changes for each T1 event. (d) MSRD, corrected by $l_v$ and scaled with $\theta_{T1}$, collapses into two separate lines as a function of $\tau_\text{eff}$. The dashed line in (a) represents unity.}
    \label{fig:6}
\end{figure}

Despite the clear differences in spatial and temporal correlations, long-time cellular movement can be universally described by persistent Brownian motion~\cite{furth1920brownsche,selmeczi2005cell}

\begin{equation}
    MSD(t) = 2v^2\tau_\text{eff}^2 \left(t/\tau_\text{eff} -1 + e^{-t/\tau_\text{eff}} \right).
    \label{eq:msd}
\end{equation}

The long-time limit of Eq.~\ref{eq:msd}, $MSD(t^*) \approx 2v^2\tau_\text{eff} \cdot t^*$, predicts the MSD for both activities without free parameters (Fig.~\ref{fig:6}(a)). After correcting $MSRD$ by $e^{-l_v}$, both activities collapse onto a single master curve (Fig.~\ref{fig:6}(b)), showing that long-time cellular movement is independent of the details of active forces.

In contrast, $\theta_\text{T1}$ does not reliably predict $MSRD(t^*)$ (Fig.~\ref{fig:6}(c)). This arises from the dependence of long-time cellular motion on persistence, as cells with higher persistence can separate by larger distances with the same number of T1 transitions (Fig.~\ref{fig:6}(c) insets, Supplementary movies 1-2).  The long-time displacement per T1 rate correlates strongly with $\tau_\text{eff}$, indicating that $\theta_\text{T1}$ alone cannot predict cellular movement, contrasting with recent studies reporting universal correlations~\cite{jain2024cell,jain2025emergent}. Even after accounting for $\tau_\text{eff}$, the correlations split into two distinct branches, implying that additional non-equilibrium features, such as the typical traveling distance between T1 transitions, are required for a universal prediction.

In this work, we investigated non-equilibrium dynamics of confluent tissues under traction or tension fluctuations and showed that cell geometry, rearrangements, and spatiotemporal dynamics strongly depend on the specific mode of activity in fluid states. These differences lead to non-universal correlations with long-time cell movement, making it infeasible to predict $MSRD(t^*)$ from either cell shape or T1 rate. In contrast, $MSRD(t^*)$ is universally captured by persistent Brownian motion, consistent with prior observations in epithelial monolayers~\cite{selmeczi2005cell,szabo2010collective,czirok2013collective}, suggesting a general description for biological tissues independent of active force details. Finally, the distinct structural and dynamical signatures between traction and tension fluctuations offer a potential method to infer the dominant active forces in tissues as an alternative approach to assess their mechanical states. Future work should explore mixed regimes with comparable active force magnitudes, together with other active processes such as cell division, apoptosis, extrusion, and insertion, to assess how universally the persistent Brownian dynamics framework can predict fluidity.

\section*{Acknowledgments}

The authors are grateful for helpful discussions with Sascha Hilgenfeldt, Cristina Marchetti, Adrien M\'{e}ry, and Selman Sakar. We thank Scientific IT \& Applications Support (SCITAS) for providing support with the high-performance computing facilities. 

\bibliographystyle{unsrt}
\bibliography{bib.bib}

@article{kim2021embryonic,
  title={Embryonic tissues as active foams},
  author={Kim, Sangwoo and Pochitaloff, Marie and Stooke-Vaughan, Georgina A and Camp{\`a}s, Otger},
  journal={Nature physics},
  volume={17},
  number={7},
  pages={859--866},
  year={2021},
  publisher={Nature Publishing Group UK London},
  doi={10.1038/s41567-021-01215-1}
}

@article{yamamoto2022non,
  title={Non-monotonic fluidization generated by fluctuating edge tensions in confluent tissues},
  author={Yamamoto, Takaki and Sussman, Daniel M and Shibata, Tatsuo and Manning, M Lisa},
  journal={Soft Matter},
  volume={18},
  number={11},
  pages={2168--2175},
  year={2022},
  publisher={Royal Society of Chemistry}
}

@article{jain2024cell,
  title = {From cell intercalation to flow, the importance of T1 transitions},
  author = {Jain, Harish P. and Voigt, Axel and Angheluta, Luiza},
  journal = {Phys. Rev. Res.},
  volume = {6},
  issue = {3},
  pages = {033176},
  numpages = {9},
  year = {2024},
  month = {Aug},
  publisher = {American Physical Society},
  doi = {10.1103/PhysRevResearch.6.033176},
  url = {https://link.aps.org/doi/10.1103/PhysRevResearch.6.033176}
}

@article{jain2025emergent,
  title={Emergent cell migration from cell shape deformations and T1 transitions},
  author={Jain, Harish P and Ho, Richard DJG and Angheluta, Luiza},
  journal={Physical Review Research},
  volume={7},
  number={3},
  pages={033020},
  year={2025},
  publisher={APS}
}

@article{mitchel2020primary,
  title={In primary airway epithelial cells, the unjamming transition is distinct from the epithelial-to-mesenchymal transition},
  author={Mitchel, Jennifer A and Das, Amit and O’Sullivan, Michael J and Stancil, Ian T and DeCamp, Stephen J and Koehler, Stephan and Oca{\~n}a, Oscar H and Butler, James P and Fredberg, Jeffrey J and Nieto, M Angela and others},
  journal={Nature communications},
  volume={11},
  number={1},
  pages={5053},
  year={2020},
  publisher={Nature Publishing Group UK London}
}

@article{kim2019simple,
  title={A simple landscape of metastable state energies for two-dimensional cellular matter},
  author={Kim, Sangwoo and Hilgenfeldt, Sascha},
  journal={Soft Matter},
  volume={15},
  number={2},
  pages={237--242},
  year={2019},
  publisher={Royal Society of Chemistry}
}

@article{furth1920brownsche,
  title={Die brownsche bewegung bei ber{\"u}cksichtigung einer persistenz der bewegungsrichtung. mit anwendungen auf die bewegung lebender infusorien},
  author={F{\"u}rth, Reinhold},
  journal={Zeitschrift f{\"u}r Physik},
  volume={2},
  number={3},
  pages={244--256},
  year={1920},
  publisher={Springer-Verlag Berlin/Heidelberg}
}

@article{selmeczi2005cell,
  title={Cell motility as persistent random motion: theories from experiments},
  author={Selmeczi, David and Mosler, Stephan and Hagedorn, Peter H and Larsen, Niels B and Flyvbjerg, Henrik},
  journal={Biophysical journal},
  volume={89},
  number={2},
  pages={912--931},
  year={2005},
  publisher={Elsevier}
}

@article{szamel2024extremely,
  title={Extremely persistent dense active fluids},
  author={Szamel, Grzegorz and Flenner, Elijah},
  journal={Soft Matter},
  volume={20},
  number={26},
  pages={5237--5244},
  year={2024},
  publisher={Royal Society of Chemistry}
}

@article{sadhukhan2024motility,
  title={Motility driven glassy dynamics in confluent epithelial monolayers},
  author={Sadhukhan, Souvik and Nandi, Manoj Kumar and Pandey, Satyam and Paoluzzi, Matteo and Dasgupta, Chandan and Gov, Nir S and Nandi, Saroj Kumar},
  journal={Soft Matter},
  volume={20},
  number={31},
  pages={6160--6175},
  year={2024},
  publisher={Royal Society of Chemistry}
}

@article{szabo2010collective,
  title={Collective cell motion in endothelial monolayers},
  author={Szab{\'o}, Andr{\'a}s and {\"U}nnep, R and M{\'e}hes, Eld and Twal, WO and Argraves, WS and Cao, Y and Czir{\'o}k, Andr{\'a}s},
  journal={Physical biology},
  volume={7},
  number={4},
  pages={046007},
  year={2010},
  publisher={IOP Publishing}
}

@article{czirok2013collective,
  title={Collective cell streams in epithelial monolayers depend on cell adhesion},
  author={Czir{\'o}k, Andr{\'a}s and Varga, Katalin and M{\'e}hes, El{\H{o}}d and Szab{\'o}, Andr{\'a}s},
  journal={New journal of physics},
  volume={15},
  number={7},
  pages={075006},
  year={2013},
  publisher={IOP Publishing}
}

@article{henkes2020dense,
  title={Dense active matter model of motion patterns in confluent cell monolayers},
  author={Henkes, Silke and Kostanjevec, Kaja and Collinson, J Martin and Sknepnek, Rastko and Bertin, Eric},
  journal={Nature communications},
  volume={11},
  number={1},
  pages={1405},
  year={2020},
  publisher={Nature Publishing Group UK London}
}

@article{caprini2020spontaneous,
  title={Spontaneous velocity alignment in motility-induced phase separation},
  author={Caprini, Lorenzo and Marini Bettolo Marconi, Umberto and Puglisi, Andrea},
  journal={Physical review letters},
  volume={124},
  number={7},
  pages={078001},
  year={2020},
  publisher={APS}
}

@article{keta2022disordered,
  title={Disordered collective motion in dense assemblies of persistent particles},
  author={Keta, Yann-Edwin and Jack, Robert L and Berthier, Ludovic},
  journal={Physical review letters},
  volume={129},
  number={4},
  pages={048002},
  year={2022},
  publisher={APS}
}

@article{barton2017active,
  title={Active vertex model for cell-resolution description of epithelial tissue mechanics},
  author={Barton, Daniel L and Henkes, Silke and Weijer, Cornelis J and Sknepnek, Rastko},
  journal={PLoS computational biology},
  volume={13},
  number={6},
  pages={e1005569},
  year={2017},
  publisher={Public Library of Science San Francisco, CA USA}
}

@article{bi2015density,
  title={A density-independent rigidity transition in biological tissues},
  author={Bi, Dapeng and Lopez, JH and Schwarz, Jennifer M and Manning, M Lisa},
  journal={Nature Physics},
  volume={11},
  number={12},
  pages={1074--1079},
  year={2015},
  publisher={Nature Publishing Group UK London}
}

@article{guillot2013mechanics,
  title={Mechanics of epithelial tissue homeostasis and morphogenesis},
  author={Guillot, Charl{\`e}ne and Lecuit, Thomas},
  journal={Science},
  volume={340},
  number={6137},
  pages={1185--1189},
  year={2013},
  publisher={American Association for the Advancement of Science}
}

@article{heisenberg2013forces,
  title={Forces in tissue morphogenesis and patterning},
  author={Heisenberg, Carl-Philipp and Bella{\"\i}che, Yohanns},
  journal={Cell},
  volume={153},
  number={5},
  pages={948--962},
  year={2013},
  publisher={Elsevier}
}

@article{mongera2018fluid,
  title={A fluid-to-solid jamming transition underlies vertebrate body axis elongation},
  author={Mongera, Alessandro and Rowghanian, Payam and Gustafson, Hannah J and Shelton, Elijah and Kealhofer, David A and Carn, Emmet K and Serwane, Friedhelm and Lucio, Adam A and Giammona, James and Camp{\`a}s, Otger},
  journal={Nature},
  volume={561},
  number={7723},
  pages={401--405},
  year={2018},
  publisher={Nature Publishing Group UK London}
}

@article{li2024relaxation,
  title={Relaxation dynamics in the self-propelled Voronoi model for epithelial monolayers},
  author={Li, Meng-Yuan and Li, Yan-Wei},
  journal={Physical Review Research},
  volume={6},
  number={3},
  pages={033209},
  year={2024},
  publisher={APS}
}

@article{mao2024mechanical,
  title={Mechanical state transitions in the regulation of tissue form and function},
  author={Mao, Yanlan and Wickstr{\"o}m, Sara A},
  journal={Nature Reviews Molecular Cell Biology},
  volume={25},
  number={8},
  pages={654--670},
  year={2024},
  publisher={Nature Publishing Group UK London}
}

@article{lenne2022sculpting,
  title={Sculpting tissues by phase transitions},
  author={Lenne, Pierre-Fran{\c{c}}ois and Trivedi, Vikas},
  journal={Nature Communications},
  volume={13},
  number={1},
  pages={664},
  year={2022},
  publisher={Nature Publishing Group UK London}
}

@article{lawson2021jamming,
  title={Jamming and arrest of cell motion in biological tissues},
  author={Lawson-Keister, Elizabeth and Manning, M Lisa},
  journal={Current Opinion in Cell Biology},
  volume={72},
  pages={146--155},
  year={2021},
  publisher={Elsevier}
}

@article{hannezo2022rigidity,
  title={Rigidity transitions in development and disease},
  author={Hannezo, Edouard and Heisenberg, Carl-Philipp},
  journal={Trends in Cell Biology},
  volume={32},
  number={5},
  pages={433--444},
  year={2022},
  publisher={Elsevier}
}

@article{angelini2011glass, 
  year     = {2011}, 
  title    = {Glass-like dynamics of collective cell migration}, 
  author   = {Angelini, Thomas E. and Hannezo, Edouard and Trepat, Xavier and Marquez, Manuel and Fredberg, Jeffrey J. and Weitz, David A.}, 
  journal  = {Proceedings of the National Academy of Sciences}, 
  issn     = {0027-8424}, 
  doi      = {10.1073/pnas.1010059108}, 
  pmid     = {21321233}, 
  pmcid    = {{PMC}3064326}, 
  pages    = {4714--4719}, 
  number   = {12}, 
  volume   = {108}
}

@article{park2015unjamming,
  title={Unjamming and cell shape in the asthmatic airway epithelium},
  author={Park, Jin-Ah and Kim, Jae Hun and Bi, Dapeng and Mitchel, Jennifer A and Qazvini, Nader Taheri and Tantisira, Kelan and Park, Chan Young and McGill, Maureen and Kim, Sae-Hoon and Gweon, Bomi and others},
  journal={Nature materials},
  volume={14},
  number={10},
  pages={1040--1048},
  year={2015},
  publisher={Nature Publishing Group UK London}
}

@article{van2009jamming, 
  year     = {2010}, 
  title    = {Jamming of soft particles: geometry, mechanics, scaling and isostaticity}, 
  author   = {Hecke, M van}, 
  journal  = {Journal of Physics: Condensed Matter}, 
  issn     = {0953-8984}, 
  doi      = {10.1088/0953-8984/22/3/033101}, 
  pmid     = {21386274}, 
  eprint   = {0911.1384}, 
  pages    = {033101}, 
  number   = {3}, 
  volume   = {22}
}

@article{o2003jamming,
  title={Jamming at zero temperature and zero applied stress: The epitome of disorder},
  author={O’hern, Corey S and Silbert, Leonardo E and Liu, Andrea J and Nagel, Sidney R},
  journal={Physical Review E},
  volume={68},
  number={1},
  pages={011306},
  year={2003},
  publisher={APS}
}

@article{petridou2019fluidization, 
  year     = {2019}, 
  title    = {Fluidization-mediated tissue spreading by mitotic cell rounding and non-canonical Wnt signalling}, 
  author   = {Petridou, Nicoletta I. and Grigolon, Silvia and Salbreux, Guillaume and Hannezo, Edouard and Heisenberg, Carl-Philipp}, 
  journal  = {Nature Cell Biology}, 
  issn     = {1465-7392}, 
  doi      = {10.1038/s41556-018-0247-4}, 
  pmid     = {30559456}, 
  pages    = {169--178}, 
  number   = {2}, 
  volume   = {21}, 
}

@article{huycke2024patterning, 
  year     = {2024}, 
  title    = {Patterning and folding of intestinal villi by active mesenchymal dewetting}, 
  author   = {Huycke, Tyler R. and Häkkinen, Teemu J. and Miyazaki, Hikaru and Srivastava, Vasudha and Barruet, Emilie and {McGinnis}, Christopher S. and Kalantari, Ali and Cornwall-Scoones, Jake and Vaka, Dedeepya and Zhu, Qin and Jo, Hyunil and Oria, Roger and Weaver, Valerie M. and {DeGrado}, William F. and Thomson, Matt and Garikipati, Krishna and Boffelli, Dario and Klein, Ophir D. and Gartner, Zev J.}, 
  journal  = {Cell}, 
  issn     = {0092-8674}, 
  doi      = {10.1016/j.cell.2024.04.039}, 
  pmid     = {38781967}, 
  pmcid    = {{PMC}11166531}, 
  pages    = {3072--3089.e20}, 
  number   = {12}, 
  volume   = {187}, 
}

@article{petridou2021rigidity,
  title={Rigidity percolation uncovers a structural basis for embryonic tissue phase transitions},
  author={Petridou, Nicoletta I and Corominas-Murtra, Bernat and Heisenberg, Carl-Philipp and Hannezo, Edouard},
  journal={Cell},
  volume={184},
  number={7},
  pages={1914--1928},
  year={2021},
  publisher={Elsevier}
}

@article{kim2024nuclear,
  title={A nuclear jamming transition in vertebrate organogenesis},
  author={Kim, Sangwoo and Amini, Rana and Yen, Shuo-Ting and Posp{\'\i}{\v{s}}il, Petr and Boutillon, Arthur and Deniz, Ilker Ali and Camp{\`a}s, Otger},
  journal={Nature materials},
  volume={23},
  number={11},
  pages={1592--1599},
  year={2024},
  publisher={Nature Publishing Group UK London}
}

@article{kim2015cell,
  title={Cell shapes and patterns as quantitative indicators of tissue stress in the plant epidermis},
  author={Kim, Sangwoo and Hilgenfeldt, Sascha},
  journal={Soft matter},
  volume={11},
  number={37},
  pages={7270--7275},
  year={2015},
  publisher={Royal Society of Chemistry}
}

@article{staple2010mechanics,
  title={Mechanics and remodelling of cell packings in epithelia},
  author={Staple, Douglas B and Farhadifar, Reza and R{\"o}per, J-C and Aigouy, Benoit and Eaton, Suzanne and J{\"u}licher, Frank},
  journal={The European Physical Journal E},
  volume={33},
  number={2},
  pages={117--127},
  year={2010},
  publisher={Springer}
}

@article{bi2016motility,
  title={Motility-driven glass and jamming transitions in biological tissues},
  author={Bi, Dapeng and Yang, Xingbo and Marchetti, M Cristina and Manning, M Lisa},
  journal={Physical Review X},
  volume={6},
  number={2},
  pages={021011},
  year={2016},
  publisher={APS}
}

@article{krajnc2020solid,
  title={Solid--fluid transition and cell sorting in epithelia with junctional tension fluctuations},
  author={Krajnc, Matej},
  journal={Soft Matter},
  volume={16},
  number={13},
  pages={3209--3215},
  year={2020},
  publisher={Royal Society of Chemistry}
}

@article{li2025fluidization,
  title={Fluidization and anomalous density fluctuations in 2D Voronoi cell tissues with pulsating activity},
  author={Li, Zhu-Qin and Lei, Qun-Li and Ma, Yu-Qiang},
  journal={Proceedings of the National Academy of Sciences},
  volume={122},
  number={10},
  pages={e2421518122},
  year={2025},
  publisher={National Academy of Sciences}
}

@article{malinverno2017endocytic, 
  year     = {2017}, 
  title    = {Endocytic reawakening of motility in jammed epithelia}, 
  author   = {Malinverno, Chiara and Corallino, Salvatore and Giavazzi, Fabio and Bergert, Martin and Li, Qingsen and Leoni, Marco and Disanza, Andrea and Frittoli, Emanuela and Oldani, Amanda and Martini, Emanuele and Lendenmann, Tobias and Deflorian, Gianluca and Beznoussenko, Galina V. and Poulikakos, Dimos and Ong, Kok Haur and Uroz, Marina and Trepat, Xavier and Parazzoli, Dario and Maiuri, Paolo and Yu, Weimiao and Ferrari, Aldo and Cerbino, Roberto and Scita, Giorgio}, 
  journal  = {Nature Materials}, 
  issn     = {1476-1122}, 
  doi      = {10.1038/nmat4848}, 
  pmid     = {28135264}, 
  pmcid    = {{PMC}5407454},  
  pages    = {587--596}, 
  number   = {5}, 
  volume   = {16}
}

@article{brugues2014forces,
  title={Forces driving epithelial wound healing},
  author={Brugu{\'e}s, Agust{\'\i} and Anon, Ester and Conte, Vito and Veldhuis, Jim H and Gupta, Mukund and Colombelli, Julien and Mu{\~n}oz, Jos{\'e} J and Brodland, G Wayne and Ladoux, Benoit and Trepat, Xavier},
  journal={Nature physics},
  volume={10},
  number={9},
  pages={683--690},
  year={2014},
  publisher={Nature Publishing Group UK London}
}

@article{alert2020physical,
  title={Physical models of collective cell migration},
  author={Alert, Ricard and Trepat, Xavier},
  journal={Annual Review of Condensed Matter Physics},
  volume={11},
  number={1},
  pages={77--101},
  year={2020},
  publisher={Annual Reviews}
}

@article{romanczuk2012active,
  title={Active Brownian particles: From individual to collective stochastic dynamics},
  author={Romanczuk, Pawel and B{\"a}r, Markus and Ebeling, Werner and Lindner, Benjamin and Schimansky-Geier, Lutz},
  journal={The European Physical Journal Special Topics},
  volume={202},
  number={1},
  pages={1--162},
  year={2012},
  publisher={Springer}
}

@article{curran2017myosin,
  title={Myosin II controls junction fluctuations to guide epithelial tissue ordering},
  author={Curran, Scott and Strandkvist, Charlotte and Bathmann, Jasper and De Gennes, Marc and Kabla, Alexandre and Salbreux, Guillaume and Baum, Buzz},
  journal={Developmental cell},
  volume={43},
  number={4},
  pages={480--492},
  year={2017},
  publisher={Elsevier}
}

@article{walck2014cell,
  title={Cell intercalation from top to bottom},
  author={Walck-Shannon, Elise and Hardin, Jeff},
  journal={Nature reviews Molecular cell biology},
  volume={15},
  number={1},
  pages={34--48},
  year={2014},
  publisher={Nature Publishing Group UK London}
}

@article{bi2014energy,
  title={Energy barriers and cell migration in densely packed tissues},
  author={Bi, Dapeng and Lopez, Jorge H and Schwarz, Jennifer M and Manning, M Lisa},
  journal={Soft matter},
  volume={10},
  number={12},
  pages={1885--1890},
  year={2014},
  publisher={Royal Society of Chemistry}
}

@article{krajnc2018fluidization,
  title={Fluidization of epithelial sheets by active cell rearrangements},
  author={Krajnc, Matej and Dasgupta, Sabyasachi and Ziherl, Primo{\v{z}} and Prost, Jacques},
  journal={Physical Review E},
  volume={98},
  number={2},
  pages={022409},
  year={2018},
  publisher={APS}
}

@article{kim2018universal,
  title={Universal features of metastable state energies in cellular matter},
  author={Kim, Sangwoo and Wang, Yiliang and Hilgenfeldt, Sascha},
  journal={Physical review letters},
  volume={120},
  number={24},
  pages={248001},
  year={2018},
  publisher={APS}
}

@article{perez2023tension,
  title={Tension remodeling regulates topological transitions in epithelial tissues},
  author={P{\'e}rez-Verdugo, Fernanda and Banerjee, Shiladitya},
  journal={PRX life},
  volume={1},
  number={2},
  pages={023006},
  year={2023},
  publisher={APS}
}

@article{saraswathibhatla2020spatiotemporal,
  title={Spatiotemporal force and motion in collective cell migration},
  author={Saraswathibhatla, Aashrith and Galles, Emmett E and Notbohm, Jacob},
  journal={Scientific Data},
  volume={7},
  number={1},
  pages={197},
  year={2020},
  publisher={Nature Publishing Group UK London}
}

@article{de2025epithelial,
  title={Epithelial layer fluidization by curvature-induced unjamming},
  author={De Marzio, Margherita and Das, Amit and Fredberg, Jeffrey J and Bi, Dapeng},
  journal={Physical review letters},
  volume={134},
  number={13},
  pages={138402},
  year={2025},
  publisher={APS}
}

@article{sussman2017cellgpu,
  title={cellGPU: Massively parallel simulations of dynamic vertex models},
  author={Sussman, Daniel M},
  journal={Computer Physics Communications},
  volume={219},
  pages={400--406},
  year={2017},
  publisher={Elsevier}
}

\end{document}